\documentclass[sigconf]{acmart}
\AtBeginDocument{%
  }

\usepackage{soul}
\usepackage{multirow}
\usepackage{colortbl}
\begin{document}


\title{What Price Fairness? Evaluating Energy -- Fairness -- Accuracy Trade-off in Recommender Systems}
%\title{Fairness at what cost? Energy-aware evaluation of recommender systems}

% The green cost of fairness-aware recommendation
% Is fairness green? Cost-aware evidence from recommender systems

\author{Abhirup Mitra}
\email{abhirup.mitra@jku.at}
\orcid{0009-0007-7171-6224}
\affiliation{%
  \institution{Johannes Kepler University Linz}
  \city{Linz}
  \country{Austria}
}

\author{Oleg Lesota}
\email{oleg.lesota@jku.at}
\orcid{0000-0002-8321-6565}
\affiliation{%
  \institution{Johannes Kepler University Linz}
  \city{Linz}
  \country{Austria}
}

\author{Antonela Tommasel}
\email{antonela.tommasel@jku.at}
\orcid{0000-0001-6091-8305}
\affiliation{%
  \institution{Johannes Kepler University Linz}
  \city{Linz}
  \country{Austria}
}
\affiliation{
  \institution{ISISTAN, CONICET-UNCPBA}
  \city{Tandil}
  \country{Argentina}
}


\begin{abstract}
Fairness-aware recommender systems aim to mitigate systematic imbalances in recommendation outcomes, including how visibility, relevance, and opportunities are distributed among users, items, and providers. 
However, these systems are usually evaluated in terms of accuracy and fairness alone, while their computational and environmental costs remain largely invisible. This omission matters because fairness interventions may affect the cost of recommendation in different ways. Training-time methods modify model optimization, post-processing methods add computation at inference time, and both may depend on the model, dataset, hardware, and deployment setting. 
We examine whether provider-side fairness in recommendation comes with a measurable green cost. We compare in-processing, graph-level reweighting and post-processing interventions across multiple models, two datasets, and two hardware settings. We measure recommendation quality, provider-side exposure, and energy consumption separately across training and  inference stages. 
Our results show that the green cost of fairness is not uniform, post-processing shifts cost to repeated serving, while in-processing and graph-level methods avoid re-ranking overhead but vary substantially across models, datasets, and hardware. 
Findings call for evaluating fairness-aware recommendation as a three-way trade-off between accuracy, fairness, and computational cost. 
% These findings call for evaluating fairness interventions not only by whether they improve exposure or preserve recommendation quality, but also 
% by where their costs arise, how they accumulate, and 
% whether the fairness gains are worth the additional energy and pipeline complexity.

\end{abstract}

\begin{CCSXML}
<ccs2012>
<concept>
<concept_id>10010583.10010662.10010673</concept_id>
<concept_desc>Hardware~Impact on the environment</concept_desc>
<concept_significance>500</concept_significance>
</concept>
<concept>
<concept_id>10002951.10003317.10003347.10003350</concept_id>
<concept_desc>Information systems~Recommender systems</concept_desc>
<concept_significance>500</concept_significance>
</concept>
</ccs2012>
\end{CCSXML}

\ccsdesc[500]{Hardware~Impact on the environment}
\ccsdesc[500]{Information systems~Recommender systems}


\maketitle

\subsubsection*{\textbf{Introduction}}
\vspace{-0.1cm}Fairness-aware recommender systems aim to mitigate systematic imbalances in recommendation outcomes, including how visibility, relevance and opportunities are distributed among users, items and providers \cite{JIN2023101906, 10.1145/3664928}. Prior work has addressed these issues through exposure-based ranking objectives \cite{DBLP:journals/tois/WuMMDL23}, fair top-$k$ re-ranking \cite{DBLP:conf/cikm/ZehlikeB0HMB17} and training-stage interventions such as upsampling and loss regularization \cite{DBLP:journals/umuai/BorattoFM21}. These approaches are commonly evaluated through the trade-off between recommendation accuracy and fairness, leaving largely invisible \textit{the computational and environmental cost of making recommender systems fairer}.

Recent work on green recommender systems has started measuring the energy and carbon footprint of recommendation pipelines, showing that model, dataset size, hardware and experimental design can affect their environmental impact \cite{DBLP:conf/recsys/SpilloFMMS23,DBLP:conf/recsys/VenteWSB24,DBLP:conf/recsys/SchodlLTS25}. However, these evaluations have mostly focused on the trade-off between quality and energy, while fairness-oriented evaluations rarely quantify the additional computational costs of fairness interventions.
We study this question for provider-side fairness, where the goal is to reduce disparities in how groups of item providers receive visibility in recommendation. Fairness interventions may act at different stages of the pipeline. Training-time methods modify optimization, while post-processing methods add computation during inference. %Their cost may therefore depend not only on the fairness objective, but also on the model, dataset and hardware.
Thus, fairness may not have a single \textit{green cost}, the same intervention may be negligible in one pipeline but substantial in another.

We take a step toward cost-aware evaluation of fairness in recommender systems. We study the trade-off between accuracy, exposure-oriented fairness and computational cost across multiple recommenders and fairness strategies. %Rather than focusing solely on fairness, w
We investigate where fairness costs arise and how they relate to the observed gains in exposure and recommendation quality.
% Our goal is not only to ask whether a recommender becomes fairer, but also whether the observed fairness gains are worth their computational and environmental cost.

\vspace{-0.15cm}\subsubsection*{\textbf{Methodology}}
\vspace{-0.1cm}We compare recommender configurations in terms of recommendation quality, exposure-oriented fairness, and energy consumption\footnote{Code and results are available at: \url{https://github.com/hcai-mms/what_price_fairness}}. All experiments are implemented in RecBole \cite{recbole}
% \footnote{\url{https://recbole.io}} 
with fixed data splits, random seed, and evaluation protocol. For each configuration, we train or load a recommender, optionally apply a fairness intervention, generate full-ranking top-$K$ lists, and measure accuracy, exposure diagnostics, and energy consumption.

\vspace{-0.0cm}\textit{Datasets and protected groups.}
We use \textit{MovieLens-10M} (ML-10M) \cite{harper2015movielens}
% \footnote{\url{https://grouplens.org/datasets/movielens/10m/}} 
and \textit{Amazon Books} (ABs) \cite{amazon}
% \footnote{\url{https://www.kaggle.com/datasets/chhavidhankhar11/amazon-books-dataset}}
, providing contrasting dense/sparse settings. Ratings are binarized by retaining interactions with rating $\geq 4$, followed by a ten-core filter and a chronological per-user 70/10/20 train/validation/test split. Since we focus on provider-side exposure, groups are defined at the item level. Motivated by work on long-tail under-exposure and provider-side fairness \cite{DBLP:conf/recsys/AbdollahpouriMB19,DBLP:journals/umuai/BorattoFM21}, we operationalize protected status through training-set popularity: the less popular half of items is treated as the protected long-tail group ($g=1$), and the more popular half as the unprotected group ($g=0$)\footnote{This median split is an experimental proxy for item-side disadvantage, not a standard definition of protected status.}.

\vspace{-0.0cm}\textit{Recommendation models.}
We evaluate BPR, NCF, MultiVAE, RecVAE, and LightGCN, covering matrix-factorization, neural, autoencoder, and graph-based recommenders. ItemKNN is added on ML-10M% as a neighbourhood-based baseline
. We use RecBole defaults, a fixed 10-epoch budget \cite{DBLP:conf/recsys/SchodlLTS25}, batch sizes of 4096/1024 for ML-10M/ABs, Recall@10 for validation.

\vspace{-0.0cm}\textit{Fairness interventions.}
We consider three provider-fairness interventions at different stages of the pipeline. As they do not optimize the same mathematical objective, they are not compared as equivalents, but rather as alternatives with the same provider-exposure-oriented goal.
First, we add a loss-level \textit{in-processing} group score-parity regularizer to trainable models \cite{DBLP:conf/nips/YaoH17,DBLP:journals/umuai/BorattoFM21}: 
$L = L_{\mathrm{base}} + \lambda(\bar{s}_{0}-\bar{s}_{1})^{2}$, 
where $\bar{s}_{g}$ is the mean predicted score for positive training items from group $g$ in a batch, and $\lambda=10^{-2}$. %This encourages the model to assign more balanced scores to protected and unprotected items during learning. The score is taken from each model's native output, without modifying the forward pass.
%: BPR and LightGCN use user--item embedding dot products, NCF uses its predicted interaction score, and MultiVAE and RecVAE use the reconstructed score of the user's interacted items. 
Second, for LightGCN \cite{DBLP:conf/sigir/0001DWLZ020}, we evaluate graph-level inverse-propensity edge re-weighting, inspired by debiased neighbour aggregation \cite{DBLP:conf/cikm/KimODL22}, and treat it as structural \textit{in-processing} because the weighted adjacency matrix is used during message passing. %Protected long-tail items therefore contribute proportionally more signal during propagation, aiming to increase their downstream exposure without altering the LightGCN architecture.
Finally, we apply FA*IR \cite{DBLP:conf/cikm/ZehlikeB0HMB17} as a model-agnostic \textit{post-processing} re-ranker over a top-500 candidate pool, enforcing a minimum protected-group proportion $\tau=0.3$ at each prefix. %If the constraint would otherwise be violated, the highest-scoring available protected item is inserted; otherwise, the highest-scoring remaining item is selected.
%This re-ranking is applied on top of each base model and, where applicable, on top of an already in-processed model.
% Together, these loss-level, graph-level, and post-processing interventions allow us to study not only whether exposure-oriented fairness improves, but also where in the pipeline its computational and environmental costs arise.

\vspace{-0.0cm}\textit{Evaluation metrics.}
Quality is measured with nDCG@10. To avoid favouring an intervention through its own objective, fairness is evaluated using out-of-objective exposure diagnostics: catalog coverage
% , exposure disparity, Shannon entropy, Gini index, long-tail exposure, 
and protected long-tail coverage \cite{10.1145/3664928}. These metrics capture broader changes in visibility and remain comparable across training-time, graph-level, and post-processing interventions.

\vspace{-0.0cm}\textit{Energy measurement.}
We use CodeCarbon \cite{benoit_courty_2024_11171501} to track energy consumption in kWh, reporting energy to avoid dependence on carbon-emission intensity factors \cite{DBLP:conf/recsys/SchodlLTS25}. Measurement interval is 15 secs.
% Following 
% work on sustainability beyond training 
We measure training and inference \cite{DBLP:conf/recsys/SchodlLTS25}. Training covers the fixed ten-epoch budget, inference covers full-item scoring, masking, and top-$K$ extraction, and re-ranking is treated as an additional serving-stage cost. Runs are repeated 3 times and averaged.

\vspace{-0.0cm}\textit{Hardware setup.}
Experiments are run on a server (Intel Core i5-4570, NVIDIA GTX 1080 Ti, Linux 6.8) and a laptop (Intel Core i9-14900HX, NVIDIA RTX 5070 Laptop GPU, Windows 11). Both use the same code, seed, split, and configuration, keeping recommendations fixed while isolating hardware-related energy differences.
% Experiments are run on two hardware setups: a \textit{server} with an Intel Core i5-4570 CPU @ 3.20 GHz and one NVIDIA GeForce GTX 1080 Ti GPU running Linux 6.8.0-117-generic (x86\_64), and a \textit{laptop} with an Intel Core i9-14900HX CPU and one NVIDIA GeForce RTX 5070 Laptop GPU running Windows 11 Home.
% Both setups use the same code, seed, data split and configuration. This paired design keeps recommendation outputs fixed while allowing energy differences to be attributed to hardware rather than to changes in the experimental protocol.
% Server: Linux 6.8.0-117-generic (x86_64) with glibc 2.39
% Laptop: Windows 11 Home.

\vspace{-0.2cm}\subsubsection*{\textbf{Experimental results}}
\vspace{-0.1cm}Table~\ref{tab:ml10m} presents the results for ML-10M\footnote{Due to space constraints, the full Amazon Books results and complete metric tables are available in the companion repository.}.

% \paragraph{Post-processing costs accumulate at inference time.}
\vspace{-0.0cm}\textit{\ul{Inference-time fairness costs accumulate with repeated serving}}
% \paragraph{\ul{Post-processing fairness costs accumulate with repeated serving.}}
% \paragraph{Energy depends on the pipeline boundary.}
% \paragraph{Serving costs change the interpretation of fairness overhead.}
%As a post-processing method, FA*IR does not add training cost directly, instead its FA*IR makes the accounting of energy costs particularly visible.
As a post-processing method, FA*IR does not add training cost directly; its additional computation is paid during inference and re-ranking. Under a serving-stage comparison, this overhead can be substantial, but it varies across models and hardware. On ML-10M, FA*IR increases inference-stage energy by $+6.5\%$ (NCF) to $+162.2\%$ (MultiVAE) on the laptop and by $+1.1\%$ (LightGCN) to $+201.6\%$ (MultiVAE) on the server. On ABs, the laptop increase ranges from $+39.2\%$ (NCF) to $+394.9\%$ (LightGCN). These differences also show that the cost of the same post-processing step is hardware- and model-sensitive.
% These relative increases should also be interpreted together with the baseline inference cost of each recommender. For energy-efficient models, even a small absolute re-ranking cost can produce a large percentage increase. Conversely, models with higher baseline inference cost may show a smaller relative FA*IR overhead, while still consuming substantially more energy overall. 
% This is visible for LightGCN on the server: compared with BPR, its plain inference energy is $3430\%$ higher on ML-10M and $27962\%$ higher on Amazon Books. Even when FA*IR re-ranking is included, LightGCN remains $1215\%$ and $6990\%$ higher, respectively. On the laptop, however, LightGCN is much closer to BPR. This reinforces that post-processing overheads should not be interpreted only as relative percentages, but alongside the absolute cost of the base recommender and the hardware on which inference is executed.
Under a single training-plus-inference view, FA*IR appears much less costly, with full-pipeline increases between $+0.1\%$ (LightGCN on ML-10M server) and $+36.2\%$ (ItemKnn on 
ML-10M server) across the measured settings. This smaller increase should not be read as evidence that FA*IR is cost-neutral. \citet{DBLP:conf/recsys/SchodlLTS25} emphasized that inference cost is often overlooked despite being repeated many times in deployed systems, and that the greener choice can change depending on the number of inference queries after training. In our setting, the same logic applies to fairness post-processing. 
FA*IR can look modest in a one-time full-pipeline evaluation while still adding a recurring serving-stage cost. 
% If re-ranking is performed once, training dominates the full-pipeline energy. If it is performed repeatedly, the additional inference cost accumulates with each recommendation request and may become the dominant factor at deployment scale. 
% Therefore, the green cost of FA*IR depends not only on the model and dataset, but also on whether we evaluate a single experimental run or a repeated deployment workload.
% Therefore, FA*IR can look modest under a one-time full-pipeline, but become expensive in high-traffic or frequently queried deployments. This makes important to specify which parts of the pipeline are included in the energy analysis. 
\textit{Fairness-aware recommenders should therefore report training and inference energy separately, and distinguish one-time full-pipeline cost from repeated deployment cost.}

% \paragraph{\ul{Training-time fairness has model-specific energy costs.}}
\vspace{-0.0cm}\textit{\ul{In-processing fairness has model-specific energy costs}}
%In-processing interventions show a model-dependent energy profile.
For several models, adding the regularizer incurs only on a moderate full-pipeline cost. On ML-10M server, the increase ranges from $+2.1\%$ (BPR) to $+5.1\%$ (NCF). On ML-10M laptop, BPR and NCF also show moderate increases.
% However, these effects are not equally stable across hardware. 
% On ML-10M laptop, BPR and NCF also show moderate increases, while MultiVAE and RecVAE show measured decreases, suggesting that short runs and hardware-level measurement noise can affect the apparent cost of small training-time changes.
Larger costs appear in specific model--dataset--hardware combinations. For LightGCN, the regularizer increases full-pipeline energy by $+46.8\%$ on ML-10M laptop, $+72.7\%$ on ML-10M server, and $+192.9\%$ on ABs laptop.
NCF also becomes costly on ABs laptop, with a $+195.8\%$ full-pipeline increase. On ABs server, however, increases are more moderate, ranging from $+5.8\%$ for BPR to $+45.2\%$ for LightGCN. 
% These results suggest that training-time fairness is not a fixed additive cost, 
% Its energy impact depends on the model architecture, dataset, and hardware, and 
% it can be small for some pipelines but substantial for others. 
The graph-level IPW intervention reinforces this point. On ML-10M, it has little or no energy effect, while on ABs it increases full-pipeline energy by $+79.4\%$ on the laptop, but only by $+0.9\%$ on the server, while also producing measurable accuracy and exposure changes.
\textit{Training-time and graph-level fairness methods therefore cannot be described as generally cheap or generally expensive, their green cost vary substantially across settings.}

\vspace{-0.0cm}\textit{\ul{Energy costs must be interpreted together with fairness gains}}
Energy increases are only meaningful in relation to the fairness gains %they produce
and the accuracy trade-offs they introduce. %FA*IR often increases long-tail exposure and coverage, but its effects on broader exposure diagnostics and ranking quality remain model-dependent. 
On ML-10M, stacking FA*IR with in-processing regularization yields large long-tail exposure gains for BPR and LightGCN% ($+13.7405$ and $+17.7920$, respectively)
, but also significantly reduces nDCG@10.
%Thus, improved long-tail exposure does not necessarily imply a favorable overall trade-off. 
On ABs, the trade-off is more model-dependent. Stacked FA*IR reduces ranking quality for BPR and RecVAE, increases protected-item fraction for MultiVAE while decreasing nDCG@10, and improves both exposure and nDCG@10 for LightGCN.
% . LightGCN provides a more favorable case, improving exposure diagnostics while also increasing nDCG@10 from $0.0110$ to $0.0122$ relative to the base model. 
These results show that intervention stage shapes the accuracy--fairness--energy trade-off, and that higher energy is not automatically justified by higher fairness. 
FA*IR offers direct control over the final top-$k$ ranking, but adds recurring serving cost and can trade long-tail gains for accuracy losses or worse broader exposure diagnostics. %, as in the Amazon Books MultiVAE and RecVAE cases. 
In-processing and graph-level interventions avoid re-ranking overhead, but remain model- and dataset-dependent. They can improve exposure and accuracy jointly, as for MultiVAE (low energy cost) and LightGCN (high energy cost) on ABs, or add energy without clear metric gains, as IPW does on ML-10M.
% In-processing and graph-level interventions avoid a separate re-ranking step, but their effects are less predictable: in some model--dataset combinations they improve exposure and accuracy jointly, while in others they increase energy without clear metric gains.
% Higher energy is therefore not automatically justified by higher fairness. On Amazon Books, for instance, FA*IR increases long-tail exposure for MultiVAE and RecVAE, but also worsens exposure disparity; for RecVAE, it additionally reduces catalog coverage and entropy while increasing Gini. Conversely, some interventions have a more favorable three-way trade-off. The in-processing regularizer improves both exposure and accuracy for MultiVAE and LightGCN on Amazon Books, although its energy cost is low for MultiVAE and much higher for LightGCN. IPW also illustrates this distinction. On ML-10M, it changes energy without producing measurable recommendation or exposure benefits, while on Amazon Books it improves both exposure diagnostics and accuracy, making the additional energy easier to justify.
\textit{Fairness gains should be considered in the context of costs in accuracy and energy.}

\vspace{-0.45cm}\subsubsection*{\textbf{Conclusions}}
% Results are indicative rather than conclusive, but they make the central point clear, fairness interventions should not be treated as cost-neutral. Their effects depend on the model, intervention stage and considered metrics, which means that fairness gains can only be properly interpreted together with their computational cost.
We evaluate pro\-vi\-der-side fairness in recommenders through a joint accuracy--fairness--cost lens. %Rather than ranking fairness strategies universally, 
Results show that the green cost of fairness is not universal, depends on the intervention type, model, dataset, hardware, and measuring scope. Interventions can look negligible under one-time full-pipeline accounting while still adding stage-specific costs under repeated deployment.
% The relevant question is not simply whether a fairness intervention increases energy, but whether the additional energy buys meaningful exposure improvements, how broadly those improvements are distributed across users, and what accuracy losses or gains accompany them.
% The study presents some limitations. We fix the fairness hyperparameters, use a single popularity-based proxy for protected items, and evaluate only two datasets with a selected set of models. Therefore, the results are not meant to be universal. In addition, energy measurements remain approximate, since CPU and memory power are estimated, and short inference or re-ranking runs are sensitive to background system noise and hardware-level variability.
Our study is limited by fixed fairness hyperparameters, a single popularity-based proxy for protected items, two datasets, models, fixed fairness parameters and approximate energy measures. 
Still, the results support making computational cost part of fairness evaluation.
Fairness-aware recommenders should be assessed by where costs arise, how they accumulate, and what trade-offs accompany exposure gains. \textit{Fairness should not be assumed computationally free, but cost should also not be used to dismiss fairness goals.}

\begin{table*}[t]
\caption{\label{tab:ml10m} Performance metrics and energy consumption for different model -- hardware -- intervention combinations on the ML-10M dataset. Absolute values shown for the backbone models, relative changes to the backbones in percent shown for the interventions. Highest relative increase in energy consumption per column highlighted. Mean model inference std below 8\%.\\\small{Additional metrics, training/inference CPU/GPU energy and inference time can be found in the repository.}}
\scalebox{0.85}{
\begin{tabular}{llrrrrrrrrrr}
\toprule
\multirow{2}{1.5cm}{\centering Model} & \multirow{2}{2.2cm}{\centering Intervention} & \multirow{2}{1cm}{\centering nDCG} & Long-tail & \multirow{2}{1.5cm}{\centering Coverage} & \multicolumn{3}{p{4cm}}{\centering Laptop energy (kWh)} & \multicolumn{3}{p{4cm}}{\centering Server energy (kWh)} \\
&  &  & coverage &  &  \multicolumn{1}{c}{Training} & \multicolumn{1}{c}{Inference} &  \multicolumn{1}{c}{Full} &  \multicolumn{1}{c}{Training} &   \multicolumn{1}{c}{Inference} &  \multicolumn{1}{c}{Full} \\
\midrule
ItemKNN & -  &  $0.0933$ & $0.0688$ & $0.2981$ & $0.0053$ & $0.0003$ & $0.0056$ & $0.0067$ & $0.0011$ & $0.0079$ \\
 &  FA*IR  &  $-1.4\%$ & $+631.7\%$ & $+48.1\%$ & $0.0\%$ & $+29.4\%$ & $+1.8\%$ & $0.0\%$ & \cellcolor{red!18}  $+253.8\%$ & $+36.2\%$ \\
\midrule
BPR & -  &  $0.0838$ & $0.0091$ & $0.156$ & $0.009$ & $0.0001$ & $0.0091$ & $0.0148$ & $0.0003$ & $0.0151$ \\
 &  FA*IR  &  $-0.5\%$ & $+785.7\%$ & $+10.3\%$ & $0.0\%$ & $+123.0\%$ & $+0.9\%$ & $0.0\%$ & $+171.3\%$ & $+3.1\%$ \\
 &  Fairness  &  $-1.1\%$ & $+611.0\%$ & $+30.3\%$ & $+9.8\%$ & $+2.3\%$ & $+9.8\%$ & $+2.1\%$ & $+0.9\%$ & $+2.1\%$ \\
 &  Fairness + FA*IR  &  $-8.9\%$ & $+1258.2\%$ & $+33.8\%$ & $+9.8\%$ & $+116.2\%$ & $+10.6\%$ & $+2.1\%$ & $+166.1\%$ & $+5.1\%$ \\
\midrule
NCF & -  &  $0.0802$ & $0.0667$ & $0.473$ & $0.0229$ & $0.0005$ & $0.0234$ & $0.0401$ & $0.0018$ & $0.0419$ \\
 &  FA*IR  &  $-21.6\%$ & $+1047.8\%$ & $+49.7\%$ & $0.0\%$ & $+6.5\%$ & $+0.1\%$ & $0.0\%$ & $+16.7\%$ & $+0.7\%$ \\
 &  Fairness  &  $+4.0\%$ & $-50.2\%$ & $-18.9\%$ & $+6.5\%$ & $-3.6\%$ & $+6.3\%$ & $+5.3\%$ & $+1.3\%$ & $+5.1\%$ \\
 &  Fairness + FA*IR  &  $-10.3\%$ & $+800.4\%$ & $+20.1\%$ & $+6.5\%$ & $+9.0\%$ & $+6.6\%$ & $+5.3\%$ & $-11.9\%$ & $+4.6\%$ \\
\midrule
MultiVAE & -  &  $0.0872$ & $0.0244$ & $0.2398$ & $0.0139$ & $0.0001$ & $0.0141$ & $0.0231$ & $0.0003$ & $0.0234$ \\
 &  FA*IR  &  $-2.5\%$ & $+459.4\%$ & $+11.2\%$ & $0.0\%$ & $+162.2\%$ & $+1.6\%$ & $0.0\%$ & $+201.6\%$ & $+2.6\%$ \\
 &  Fairness  &  $0.0\%$ & $-18.0\%$ & $-0.4\%$ & $-34.3\%$ & $+0.7\%$ & $-33.9\%$ & $+2.9\%$ & $+0.8\%$ & $+2.9\%$ \\
 &  Fairness + FA*IR  &  $-2.5\%$ & $+441.0\%$ & $+9.8\%$ & $-34.3\%$ &  \cellcolor{red!18} $+164.3\%$ & $-32.3\%$ & $+2.9\%$ & $+156.5\%$ & $+4.8\%$ \\
\midrule
RecVAE & -  &  $0.091$ & $0.0007$ & $0.2042$ & $0.0126$ & $0.0002$ & $0.0128$ & $0.0313$ & $0.0003$ & $0.0316$ \\
 &  FA*IR  &  $-0.3\%$ & $+3914.3\%$ & $+2.9\%$ & $0.0\%$ & $+135.9\%$ & $+1.7\%$ & $0.0\%$ & $+162.4\%$ & $+1.5\%$ \\
 &  Fairness  &  $-2.1\%$ & $0.0\%$ & $+5.4\%$ & $-34.3\%$ & $-0.6\%$ & $-33.9\%$ & $+3.8\%$ & $+1.0\%$ & $+3.7\%$ \\
 &  Fairness + FA*IR  &  $-2.4\%$ & $+4200.0\%$ & $+8.1\%$ & $-34.3\%$ & $+141.2\%$ & $-32.1\%$ & $+3.8\%$ & $+157.0\%$ & $+5.2\%$ \\
\midrule
LightGCN & -  &  $0.0749$ & $0.0102$ & $0.073$ & $0.0372$ & $0.0001$ & $0.0372$ & $0.0915$ & $0.0097$ & $0.1012$ \\
 &  FA*IR  &  $-0.1\%$ & $+175.5\%$ & $+5.1\%$ & $0.0\%$ & $+117.1\%$ & $+0.2\%$ & $0.0\%$ & $+1.1\%$ & $+0.1\%$ \\
 &  Fairness  &  $+1.5\%$ & $+168.6\%$ & $+44.9\%$ & \cellcolor{red!18} $+46.8\%$ & $-15.1\%$ & $+46.7\%$ &  \cellcolor{red!18} $+80.5\%$ & $-1.2\%$ & $+72.7\%$ \\
 &  Fairness + FA*IR  &  $-9.2\%$ & $+451.0\%$ & $+46.2\%$ &  \cellcolor{red!18} $+46.8\%$ & $+117.8\%$ &  \cellcolor{red!18} $+46.9\%$ & \cellcolor{red!18} $+80.5\%$ & $+6.5\%$ &  \cellcolor{red!18} $+73.4\%$ \\
 &  IPW  &  $0.0\%$ & $0.0\%$ & $0.0\%$ & $+8.4\%$ & $-14.7\%$ & $+8.4\%$ & $+0.1\%$ & $-1.5\%$ & $-0.0\%$ \\
 &  IPW + FA*IR  &  $-0.1\%$ & $+174.5\%$ & $+5.1\%$ & $+8.4\%$ & $+137.5\%$ & $+8.7\%$ & $+0.1\%$ & $+6.5\%$ & $+0.7\%$ \\
\bottomrule
\end{tabular}
}
\end{table*}

\noindent \textbf{Acknowledgments. }This research was funded in whole or in part by the Austrian Science Fund (FWF): \href{https://doi.org/10.55776/COE12}{10.55776/COE12}. %, and %\href{https://doi.org/10.55776/DFH23}{10.55776/DFH23}, and \href{https://doi.org/10.55776/P36413}{10.55776/P36413}.
%and received financial support from the State of Upper Austria and the Federal Ministry of Education, Science, and Research, through grant LIT-2024-13-SEE-111.
This work received financial support from the State of Upper Austria and the Federal Ministry of Education, Science, and Research, through grant LIT-2024-13-SEE-111.

\bibliographystyle{ACM-Reference-Format}
\bibliography{references}

\end{document}